# Spin glass behavior in amorphous $Cr_2Ge_2Te_6$ phase-change alloy

Xiaozhe Wang[1#], Suyang Sun[1#], Jiang-Jing Wang[1*], Shuang Li[1], Jian Zhou[1], Oktay Aktas[2], Ming Xu[3], Volker L. Deringer[4], Riccardo Mazzarello[5], En Ma[1], Wei Zhang[1*]

[1]Center for Alloy Innovation and Design (CAID), State Key Laboratory for Mechanical Behavior of Materials, Xi'an Jiaotong University, Xi'an 710049, China
[2]State Key Laboratory for Mechanical Behavior of Materials, Xi'an Jiaotong University, Xi'an 710049, China
[3] Wuhan National Laboratory for Optoelectronics, School of Integrated Circuits, Huazhong University of Science and Technology, Wuhan 430074, China
[4]Department of Chemistry, Inorganic Chemistry Laboratory, University of Oxford, Oxford OX1 3QR, UK
[5]Department of Physics, Sapienza University of Rome, Rome, 00185, Italy

[#]These authors contributed equally to this work.

*Emails: j.wang@mail.xjtu.edu.cn, wzhang0@mail.xjtu.edu.cn

## Abstract
The layered crystal structure of $Cr_2Ge_2Te_6$ shows ferromagnetic ordering at the two-dimensional limit, which holds promise for spintronic applications. However, external voltage pulses can trigger amorphization of the material in nanoscale electronic devices, and it is unclear whether the loss of structural ordering leads to a change in magnetic properties. Here, we demonstrate that $Cr_2Ge_2Te_6$ preserves the spin-polarized nature in the amorphous phase, but undergoes a magnetic transition to a spin glass state below 20 K. Quantum-mechanical computations reveal the microscopic origin of this transition in spin configuration: it is due to strong distortions of the Cr−Te−Cr bonds, connecting chromium-centered octahedra, and to the overall increase in disorder upon amorphization. The tunable magnetic properties of $Cr_2Ge_2Te_6$ could be exploited for multifunctional, magnetic phase-change devices that switch between crystalline and amorphous states.

# 1. Introduction

Artificial intelligence and big data analytics require continued advances in data storage and processing. Emerging electronic materials, such as phase-change materials (PCM)[1-10], resistive-switching oxides[11], spintronic materials[12], as well as two-dimensional (2D) materials[13], are being heavily investigated for the development of non-volatile memory and neuro-inspired computing, which hold promises to substantially improve the computing and power efficiencies of electronic devices[14]. Recently, the 2D van der Waals (vdW) crystal $Cr_2Ge_2Te_6$ ("CrGT" in the following) has attracted much attention, since it exhibits both long-range ferromagnetic (FM) ordering with a negligible coercivity below the Curie temperature (~66 K) and semiconducting characteristics[15]. Few-layer samples of crystalline CrGT also show other interesting properties, including electric-field controlled magnetism[16-19], pressure-induced electronic and structural transitions[20-22], high thermoelectric performance[23], and an ultrasensitive photoresponse[24]. These features make CrGT a promising candidate for future spintronic and nanoelectronic applications in low dimensions[25].

However, CrGT can readily lose its long-range structural order in nanoscale electronic devices: a moderate voltage pulse of 3 V over 100 ns is already sufficient to trigger amorphization of CrGT[26], which would significantly alter its magnetic properties. By applying a weaker pulse of 1.6 V over 30 ns, amorphous CrGT can be crystallized again[27]. This tunable phase transition capacity makes CrGT also an interesting candidate for PCM applications. In contrast with conventional PCMs, such as $Ge_2Sb_2Te_5$ ("GST")[1], CrGT shows an inverse contrast in electrical resistance with a low-resistance state in the amorphous phase, but a high-resistance state in the crystalline phase[26-28]. Regarding the optical contrast, CrGT shows an increase in optical reflectivity upon crystallization[29], similar to GST. Chromium has also been used as an alloying element to enhance the amorphous-phase stability of conventional PCMs[30-33], but did not qualitatively affect the resistance contrast. Incorporating 3$d$ transition metals in GST also leads to a magnetic contrast between the two phases, giving rise to a magnetic PCM (mPCM)[34]. In particular, Fe-doped GST exhibits ferromagnetic order in both crystalline and amorphous phase, and a sizable change in magnetic moment over ~30% is achieved upon crystallization – in addition to the large change in electrical resistance and optical reflectivity[34, 35]. Cr-doped GST and GeTe were also predicted to be mPCMs[36-38]. It remains elusive why the resistivity contrast upon phase transition is reversed in CrGT and whether this 2D vdW semiconductor can also be exploited for mPCM applications.

From a more fundamental perspective, it is intriguing to explore whether the structural transition into a glass also leads to a change in the spin configuration of CrGT, viz. from ferromagnetic order to a spin glass[39]. Amorphous magnetic materials, in particular, bulk metallic glasses[40-43] and their oxides[44-46], have been heavily investigated recently. Given the absence of structural anisotropy and crystal defects, high solubility for doping and alloying, and the compatibility with the semiconductor manufacturing of film deposition, amorphous magnetic materials hold great promises for various electronic and magnetic applications[40-46]. In the present work, we characterize the magnetic properties of amorphous CrGT by combining experiments on as-deposited thin films and computational modeling. The disordered nature and homogenous elemental distribution of the films are confirmed by transmission electron microscopy (TEM) and energy-dispersive X-ray (EDX) measurements. We demonstrate the spin-glass nature of the system via thorough magnetic characterization. *Ab initio* simulations based on density functional theory (DFT) show that amorphous CrGT contains similar octahedral motifs as the crystalline phase – albeit the presence of angular disorder and chemical bond distortions weakens the magnetic order, and gives rise to the coexistence of ferromagnetic and antiferromagnetic (AFM) coupling, consistent with our experimental observations.

## 2. Results and Discussion

### 2.1 Materials synthesis and characterizations.

**Figure 1**a shows the structure of crystalline (c-) CrGT, which contains three atomic slabs and three vdW gaps in a hexagonal unit cell. In each atomic slab, two Cr atoms form octahedral motifs with six Te atoms, and the two Ge atoms form intergrown tetrahedra with the six Te atoms, as highlighted by orange and blue polyhedra. Such a layered structure is clearly identified by atomic imaging experiments via spherical aberration corrected (Cs-corrected) scanning transmission electron microscopy (STEM) - high angle annular dark field (HAADF). Figure 1b shows the HAADF image viewed along the [120] direction measured for the single-crystal sample. The DFT-optimized lattice parameters are *a* = 6.87 Å and *c* = 20.36 Å, in good agreement with the experimental ones from XRD, 6.84 Å and 20.51 Å, respectively (Figure 1c). The single-crystal sample, prepared by via chemical vapor transport, exhibits prominent (003*n*) diffraction peaks, indicating high sample quality. Our Raman spectroscopy experiments of the single crystal sample (figure S1) are also in good agreement with literature data[47]. To prepare amorphous samples, we deposited CrGT thin films of ~200 nm thickness by sputtering a stoichiometric $Cr_2Ge_2Te_6$ target. The as-deposited CrGT thin film showed no visible no well-defined diffraction peaks, confirming that the initial phase was fully amorphous. Despite its highly disordered internal structure, the as-deposited amorphous (a-) CrGT thin film showed a relatively low resistivity value. As displayed in Figure 1d, the thin film crystallized upon *in situ* heating to 380 °C, reaching a higher resistivity value at room temperature. The corresponding XRD pattern shows that the crystallized CrGT film was polycrystalline, with fine grains as indicated by the obviously broadened peaks. These measurements and calculations are consistent with previous literature data[27, 48-51]. Details of sample preparation, characterization, and computations are given in the Methods section.

The direct-current (DC) magnetic measurements were carried out using a superconducting quantum interference device (SQUID) magnetometer with a maximum field of 70 kOe. Figure 1e shows the temperature-dependent magnetization (M-T) of the single-crystal sample. The zero-field-cooled (ZFC) and field-cooled (FC) magnetization curves were measured from 2 K to 150 K in an applied field of 2 kOe perpendicular to the crystal basal plane (H // c). These two curves were fitted by the Curie–Weiss law, $\chi = c/(T - \theta)$, where c is the Curie constant and θ is the Curie–Weiss temperature. The obtained fitting parameter of θ = 67.68 K is positive (figure S2), indicating that c-CrGT is stabilized by ferromagnetic exchange couplings.

The Curie temperature ($T_c$) was estimated to be 64.2 K from the minimum of the d$M$/d$T$, consistent with Ref. [15]. We also obtained the field-dependent magnetization (M-H) measured at 2 K and 150 K with an increasing field from –70 kOe to 70 kOe (Figure 1f). The linear magnetic hysteresis curve at 2 K shows that c-CrGT is a soft FM material (coercivity ~64 Oe) at this temperature with a saturation field of 5000 Oe, while the M-H curve at 150 K corresponds to normal paramagnetic behavior.

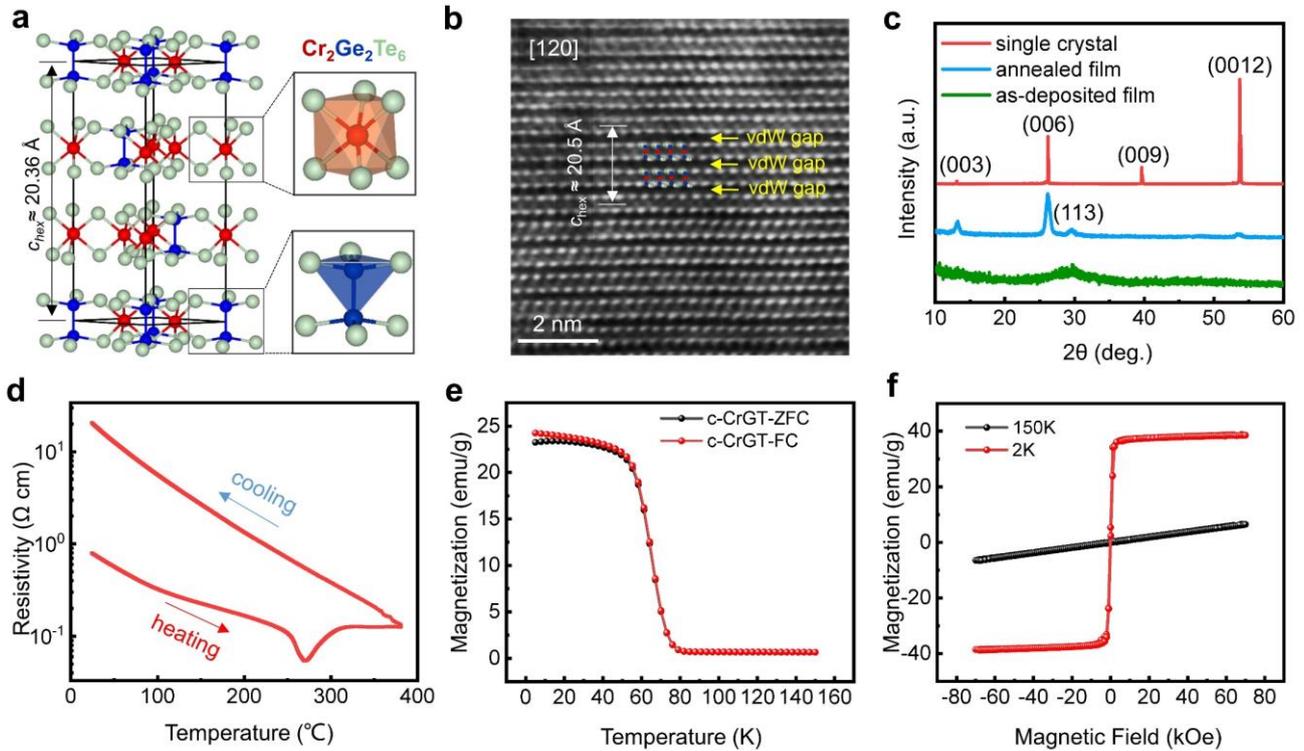

**Figure 1. Structure magnetism of crystalline CrGT. a** The atomic structure after DFT optimization. **b** The HAADF image of the single-crystal sample. **c** The XRD pattern of the single-crystal sample, in comparison with the thin films (as-deposited and crystallized after annealing at 380 °C for 10 min). **d** Temperature dependence of the resistivity (R-T) of the ~200 nm-thick CrGT film heated with a rate of 10 °C /min. The sample was held at 380 °C for 10 min and was then cooled to room temperature. **e** M-T curves in ZFC and FC measured for the single-crystal sample under an applied DC magnetic field of 2 kOe. **f** Field-dependent magnetization measured for the single-crystal sample at 2 K and 150 K with a maximum field value of 70 kOe.

## 2.2 Magnetic properties of amorphous CrGT.

It is much more challenging to conduct the same magnetic measurement for the amorphous phase, because it is difficult to produce macroscopic amorphous CrGT samples directly and the thin-film samples showed too weak signal for magnetic characterization. We have deposited thin films from ~200 nm up to ~1 μm, but mostly observed the signal from the substrate. One notes that even though weak magnetic signals of thin films can be detected by magneto-optic Kerr effect experiments[16, 52], here we chose a different route that allows similar magnetic characterizations via SQUID as done for the single crystalline sample. Specifically, we adopted the procedure shown in **Figure 2**a to produce powder samples of amorphous CrGT. A similar approach has been adopted to prepare various amorphous PCM powder samples in Ref. [53]. Firstly, we deposited ~1 μm CrGT thin film on an oil-based ink coated $SiO_2$/Si substrate. Then we immersed the sample in acetone over 12 hours to exfoliate the CrGT thin film from the substrate. after the cleaning and drying process, powdered CrGT samples were obtained. This process was repeated multiple times to acquire a sufficient amount of powder (~5.2 mg), comparable to the mass of the single-crystal sample (~5.5 mg). The whole process was done at room temperature to prevent crystallization. We performed XRD measurements on the CrGT powders, and observed no diffraction peaks. Our TEM measurements also revealed the highly disordered nature of both the as-deposited thin film (figure S3) and the final powder sample (Figure 2b), given the absence of crystallites in bright-field images and the dim halos in the corresponding selected area electron diffraction (SEAD) pattern. The corresponding EDX measurements confirmed homogenous elemental distribution in the amorphous CrGT sample, and no phase separation, e.g., Cr precipitates, could be observed at the scale of tens of nanometers. The Raman vibration modes of the amorphous thin film sample and powder sample are highly consistent (figure S1).

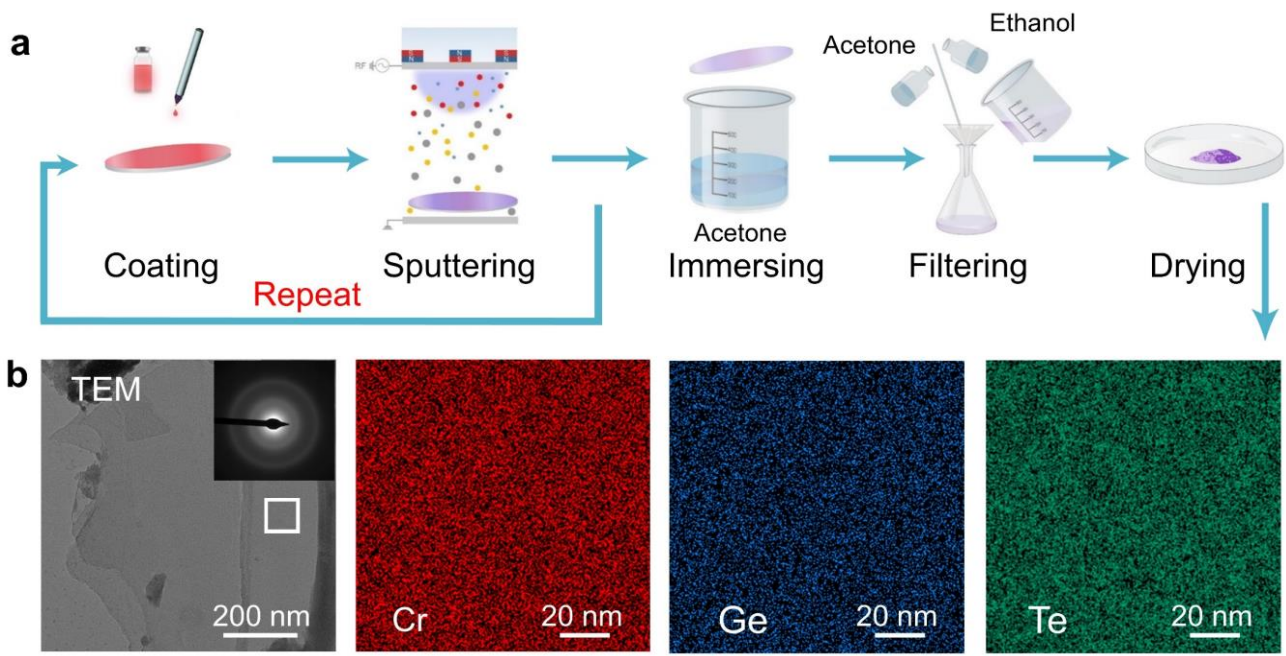

**Figure 2. Preparation of a-CrGT sample and TEM characterization. a** The synthesis process of a-CrGT powders, including coating, sputtering, immersing, filtering and drying. **b** The bright-field TEM image and the SAED pattern of the powder sample, and the corresponding EDX elemental mapping.

**Figure 3**a shows a photograph of the a-CrGT powders, which were encapsulated for magnetic measurements. The M-H curves of a-CrGT at 2 K and 150 K are shown in Figure 3b. Similar to c-CrGT, a linear M-H curve was observed at 150 K, corresponding to a paramagnetic state. At 2 K, the M-H curve showed a clear hysteresis loop, indicating the presence of FM order. Moreover, the M-H curve was almost linear at high DC magnetic fields above 20 kOe, and remained unsaturated after the maximum magnetic field of 70 kOe was reached. Such behavior could be attributed to the existence of collinear AFM fluctuations or to the random distribution of magnetic moments. Figure 3c shows a zoomed-in M-H curve of a-CrGT measured at 2K, where the hysteresis loop can be better perceived. A much larger coercivity value of ~1150 Oe with respect to c-CrGT ~64 Oe was identified in the amorphous

phase, which confirms the robustness of magnetic behavior in a-CrGT. Taking into account the net magnetization of a-CrGT at low temperatures, i.e. ~0.52 emu/g at 2 K, under a moderate external field 2000 Oe (Fig. 3d), we can conclude that a-CrGT shows a much more complex magnetic behavior with possible coexistence of both FM and AFM couplings as compared to its crystalline FM counterpart.

**2.3 Observation of spin glass behavior.**

To determine whether a-CrGT is a spin glass, we carried out systematic temperature-dependent magnetic characterizations as a function of applied DC magnetic field. Both the ZFC and FC M-T curves were recorded in the temperature range 2K to 150 K under magnetic fields with intensity ranging from 2000 Oe to 500 Oe. The magnetic behavior under various magnetic fields are shown in Figure 3d-f. At low temperatures, a-CrGT showed clear magnetic moments, while a rapid decay in magnetization was observed as the temperature approached the temperature range of 20–30 K (depending on the applied magnetic field). At higher temperatures, the M-T curve became flat, indicating a transition to a paramagnetic state. The ZFC and FC curves nearly overlapped under 2000 Oe. By fitting the susceptibility of the two M-T curves using the Curie–Weiss law, a negative θ value (–15.22 K) was obtained (figure S2), indicating the presence of AFM interactions. Under a weaker magnetic field (1000 Oe), a clear splitting in the ZFC and FC curves was observed (Figure 3e). A visible drop in magnetization was captured below ~7 K, which is regarded as the freezing temperature ($T_f$) of the spin glass. By further reducing the applied magnetic field to 500 Oe, the $T_f$ value was slightly increased to ~8 K (Figure 3f). This observation is consistent with the typical spin glass behavior in that the frozen spin glass state sets in at higher temperature with reduced applied magnetic field.

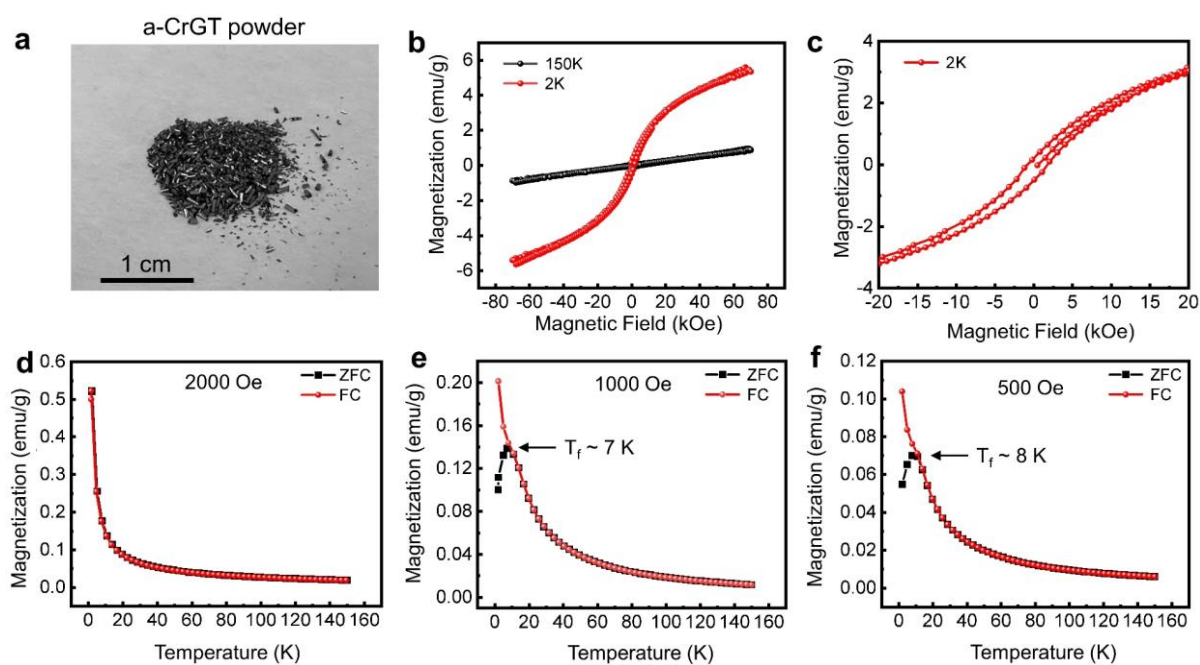

**Figure 3 The DC magnetic measurements of a-CrGT. a** Photograph of the a-CrGT powder sample. **b** The magnetization measured for a-CrGT at 2 K and 150 K with a maximum field of 70 kOe. **c** The zoomed-in M-H curve measured at 2 K. **d-f** The ZFC and FC M-T curves measured under an applied field of 2000 Oe, 1000 Oe and 500 Oe.

To gain further understanding of the magnetic behavior of a-CrGT, we carried out alternating-current (AC) magnetic susceptibility experiments. **Figure 4**a shows the real part of the magnetic susceptibility χ' measured at various frequencies ranging from 10 Hz to 500 Hz between 6 K and 19 K with an AC magnetic field of 7.5 Oe. As the frequency increases, the position of these peaks shifts to higher temperatures, while the peak height decreases, indicating typical spin glass behavior. The frequency dependence of the peak shift can be determined by

$$\Phi = \Delta T_p / (T_p \Delta log_{10} f),$$

where

$$\tau = \tau_0 [\frac{T_p}{T_0} - 1]^{-zv}$$

where the $T_p$ stands for the finite static freezing temperature, $\tau_0$ represents the characteristic flipping time of the magnetic moments, and *zv* is the dynamical critical exponent. According to previous investigations for conventional spin glasses, *zv* is usually found between 4 and 13, and $\tau_0$ ~$10^{-10}$–$10^{-13}$ s[54]. The parameters obtained for amorphous CrGT are: $T_0$ = 8.43 K, *zv* = 9.15, and $\tau_0$ = 8.9×$10^{-11}$ s (figure inset), confirming its spin glass behavior. Further evidence for a spin glass phase in a-CrGT was found by performing time-dependent remanent magnetization $M_r$ measurements. After cooling the sample down to 2 K in the ZFC mode, a DC magnetic field of 10 kOe was applied for 5 minutes. After the removal of the magnetic field, a slow decay in isothermal remanent magnetization was observed, as shown in Figure 4b. By plotting the time dependent magnetization in logarithmic scale as log {-d/d*t* [ln$M_r$]} versus log {*t*}, the slope can be fitted as a straight line (Figure 4b inset) with –*n* = –0.66. Altogether, these observations provide compelling evidence for a complex magnetic structure in a-CrGT. In order to confirm the stability of these magnetic features in a-CrGT, we have performed the same set of magnetic measurements 35 days after the powder sample was produced, and obtained very similar M-T curves (figure S4).

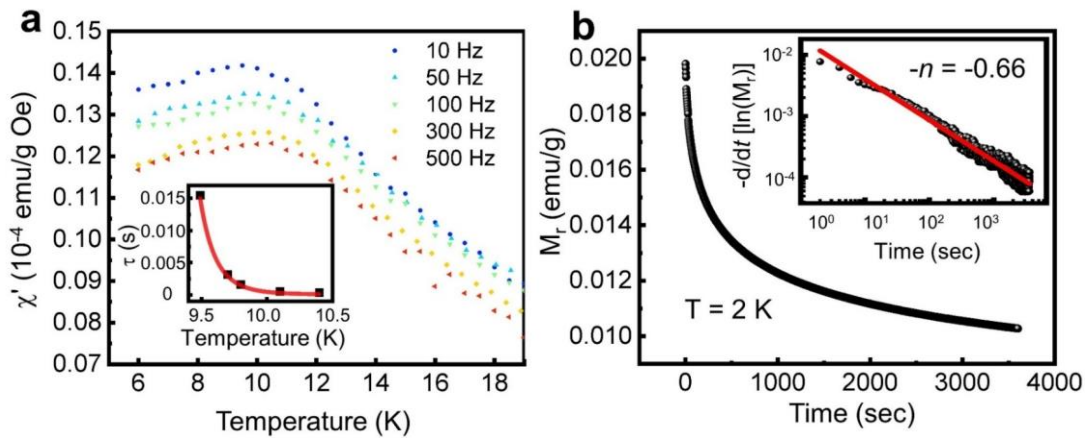

**Fig. 4 The AC magnetic susceptibility measurements and isothermal remanent magnetization of a-CrGT. a** Temperature dependence of the real part of the AC susceptibility measured at frequencies f =10, 50, 100, 300, and 500 Hz with an AC magnetic field of 7.5 Oe. Inset: fitting of the relaxation time against the power law. **b** Time dependence of the isothermal remanent magnetization $M_r$ after removal of a strong DC magnetic field of 10 kOe (T = 2 K). Inset: change of $M_r$ with time on a logarithmic scale.

## 2.4 *Ab initio* modeling of amorphous CrGT.

To shed light on the experimental findings, we generated a-CrGT models through melt-quench AIMD simulations. Specifically, a cubic cell model containing a total of 180 atoms was heated above 2000 K for randomization, quenched to 1200 K and held at that temperature for 30 ps. The liquid phase was then quenched to 300 K with a cooling rate of 12.5 K ps$^{-1}$, and was kept at 300 K for 30 ps. Subsequently, this amorphous model was cooled down to 10 K for structural data collection (30 ps at 10 K) and to 0 K for electronic structure calculations. Both non-spin-polarized and spin-polarized (with no restriction on spin direction) configurations were considered during the melt-quench runs. For each set of calculations, three amorphous models with independent thermal history were created to improve the statistics. As the difference in mass density between amorphous and crystalline CrGT is very small[27, 55], we fixed cell size at 17.1×17.1×17.1 Å$^3$, corresponding to the density of the crystalline model. Indeed, the amorphous models showed low stress values of a few kbar. The atomic structure and element-resolved radial distribution functions (RDFs) of the spin-polarized (sp-) and non-spin-polarized (nsp-) a-CrGT models annealed at 10 K are shown in **Figure 5**a and figure S5. The Cr, Ge and Te atoms are rendered in red, blue, and green spheres, respectively.

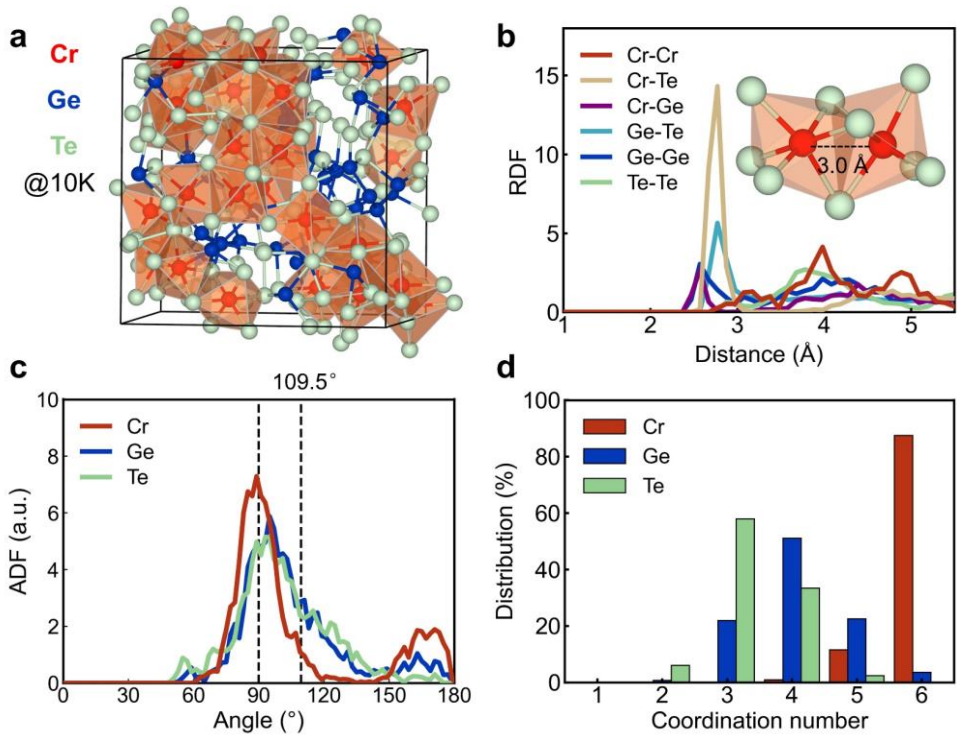

**Figure 5**. *Ab initio* simulation models of a-CrGT. **a** The relaxed structural model of *a*-CrGT in a cubic supercell with an edge length of 17.10 Å, annealed at 10 K. The Cr, Ge and Te atoms are rendered as red, blue, and green spheres, respectively. **b** Radial distribution functions obtained from the spin-polarized AIMD simulations. **c** Angular distribution function and **d** coordination number distribution of the spin-polarized amorphous models.

Clearly, there exists a major difference in the glass structure depending on whether or not spin-polarization is included. In the nsp-model, the Cr−Cr RDF shows a dominant peak below 2.0 Å, corresponding to compact Cr clusters[55]. However, these short Cr−Cr bonds appear to be a spurious effect of the nsp-approximation: besides being much shorter than Cr−Cr bonds in bcc Cr crystal (~2.5 Å) by around 20%, there is no experimental evidence for their existence. In fact, the Cr−Cr bonds determined by extended X-ray absorption fine structure (EXAFS) measurements at 10 K are much longer, on the order of ~3.0 Å in a-CrGT[56]. This measurement agrees with our sp-model, which shows a visible Cr−Cr pair correlation at around 3.0 Å but no peaks at 2.0 Å. The inset in the left panel of Figure 5b highlights this atomic contact, which corresponds to the interatomic distance between two adjacent octahedral motif centers. As displayed in Figure 5a and figure S5 (with two additional sp-models), we also found a tendency for Cr clustering. However, the cluster units are Cr-centered octahedral motifs in the sp-models, rather than Cr atoms as in the nsp-models. Spin polarization also leads to slightly longer Cr−Te bonds. The RDF first peak position shifts from ~2.6 Å (nsp-models) to ~2.8 Å (sp-models). Regarding Ge−Te and Ge−Ge contacts, no obvious difference between sp- and nsp-results can be found. The peak positions remain ~2.8 Å (Ge−Te) and ~2.6 Å (Ge−Ge) under both conditions. All these peak positions are consistent with the EXAFS fitting results[56]. These structural features do not change appreciably for larger model size (figure S6). These results indicate that inclusion of spin polarization is essential for the correct description of the structural and electronic properties of a-CrGT.

The angular distribution function (ADF) and the distribution of coordination numbers (CNs) of the sp-models are presented in Figure 5c and 4d. The cutoffs for interatomic distances were determined using a chemical bonding analysis scheme developed for amorphous PCMs[57, 58] (see figure S7). The bond angles show a clear peak close to 90° for all Cr, Ge and Te atoms, and a secondary peak for Cr and Ge atoms close to 180°, indicating the abundance of octahedral motifs. The vast majority (~87%) of Cr atoms are six-fold coordinated, and the others form defective octahedral motifs with five-fold coordination (~13%). In regard to the Cr atoms with CN = 6, the majority of them (~63% of all) are bonded with six Te atoms but no Ge, forming $Cr[Te_6]$ octahedra; ~23% of Cr atoms connected with one Ge atom, i.e., are in $Cr[Ge_1Te_5]$ octahedra, and the rest of them are bonded with less than six neighbors. Thus, the local order of a-CrGT is overall similar to that of the crystalline counterpart, where all Cr atoms form $Cr[Te_6]$ motifs. In contrast, for the artificial nsp-case, the CN for Cr atoms is largely increased and the associated chemical bonds are much more distorted due to Cr clustering. Even if finite magnetic moments are initially assigned to the Cr atoms in the nsp-models, they drop to zero in a self-consistent calculation due to the close atomic packing around Cr atoms[55]. The structural details of a- and c-CrGT models are summarized in Table 1.

**Table 1. Structural data of CrGT and three chromium tellurides by *ab initio* calculations.** For a-CrGT, the Cr−Te and Cr−Cr interatomic distances and the Cr−Te−Cr bond angle are referred to the peak position of the corresponding RDF and ADF curves.

|  | a-CrGT | c-CrGT | h-CrTe | 1T-CrTe$_2$ | m-CrTe$_3$ |
| --- | --- | --- | --- | --- | --- |
|  | spin glass | FM | FM | FM | AFM |

| | | | | | | |
|---|---|---|---|---|---|---|
| Cr−Te (Å) | | 2.77 | 2.77 | 2.79 | 2.68 | 2.66 ~ 2.75 |
| Cr−Cr (Å) | fs | 3.17 | / | 3.05 | / | / |
| | es | 3.98 | 3.97 | 4.05 | 3.79 | 3.50 |
| | cs | 4.89 | / | / | / | 4.6 |
| ∠Cr−Te−Cr (°) | fs | 69 | / | 66 | / | / |
| | es | 95 | 92 | 92 | 90 | 80 |
| | cs | 123 | / | / | / | 115 |
| Number of atoms | | 180 | 30 | 4 | 3 | 32 |
| Lattice parameters | a (Å) | 17.10 | 6.87 | 4.05 | 3.79 | 10.92 |
| | b (Å) | 17.10 | 6.87 | 4.05 | 3.79 | 11.49 |
| | c (Å) | 17.10 | 20.36 | 6.12 | 5.96 | 7.91 |
| | α (°) | 90 | 90 | 90 | 90 | 117.8 |
| | β (°) | 90 | 90 | 90 | 90 | 90 |
| | γ (°) | 90 | 120 | 120 | 120 | 90 |

## 2.5 Chemical bonding and electronic structure.

We carried out electronic structure calculations to understand the magnetic properties of a-CrGT in depth. **Figure 6** presents the computed density of states (DOS) and the crystal orbital Hamilton population (COHP)[59] chemical-bonding analysis of both crystalline and amorphous CrGT in various magnetic configurations. Chromium atoms with spin up (down) configurations are marked by red (blue) arrows, respectively. In the hypothetical non-magnetic (NM) case, c-CrGT shows a major DOS peak at the Fermi level ($E_F$), causing a strong antibonding interaction (−COHP ≪ 0) that destabilizes the system. With the inclusion of spin polarization, a narrow band gap is opened for c-CrGT regardless of the magnetic configuration, for which we considered different options: FM or AFM (Néel, zigzag or stripy). It has been demonstrated that c-CrGT is an intrinsic FM semiconductor[15] with a band gap of ~0.38 eV, as measured by angle-resolved photoemission spectroscopy[49]. Our computation for c-CrGT in the FM configuration yields a similar gap of 0.33 eV, and shows drastically reduced antibonding interaction in the vicinity of $E_F$ as compared to the hypothetical NM configuration. For the occupied bands, an antibonding peak is still present around −1.0 eV for the spin up majority states, while bonding interactions are mostly found for the spin down states. The computed local magnetic moments for Cr, Ge and Te atoms are 3.16, 0.05 and −0.10 $\mu_B$, respectively. The total energy of the FM configuration is ~285 meV/atom lower than that of the NM configuration, making it clearly more favorable. We also evaluated the impact of AFM couplings on the electronic structure and chemical bonding of c-CrGT. The three considered AFM configurations, namely, Néel, zigzag and stripy, show a higher total energy than that of FM by 9.6, 9.8 and 5.3 meV/atom, respectively. Across these AFM configurations, the average local magnetic moment for Cr atoms is 3.04 $\mu_B$, while those of Ge and Te atoms are marginal. The AFM couplings also open a narrow gap and suppress antibonding interactions at $E_F$, while their DOS and −COHP profiles almost overlap for the spin up and spin down states.

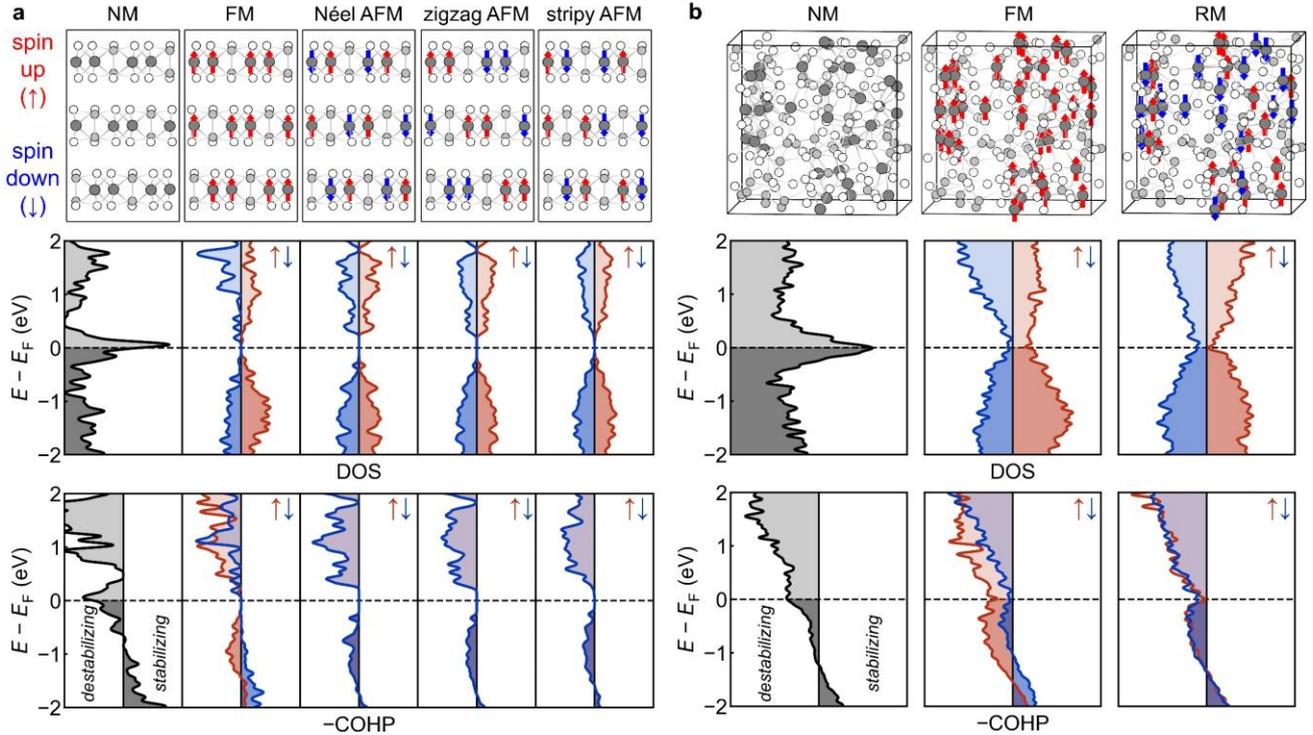

**Figure 6. Magnetic and bonding analyses. a** The atomic and magnetic structure, DOS and –COHP of c-CrGT in NM, FM, and three AFM configurations. **b** The atomic and magnetic structure, DOS and –COHP of one a-CrGT model in NM, FM and RM configurations. The Cr, Ge and Te atoms are represented by black, gray, and white spheres, respectively, with spin up and spin down moments being highlighted by red and blue arrows.

The explanation of the onset of FM ordering through (partial) removal of antibonding interactions in crystalline CrGT is fully in line with the work of Landrum and Dronskowski on transition metals[60], and we note in passing that such approaches have been successfully applied to more complex intermetallic phases[61, 62]. Here, we applied such an approach to understand magnetism in an amorphous material. We used the same atomic arrangement to study different magnetic configurations of a-CrGT, and the computations were repeated for two other amorphous models with independent thermal history. As shown in Figure 6b and figure S8, a prominent DOS peak at $E_F$ can also be found in the amorphous phase, indicating strongly destabilizing interactions in NM a-CrGT. When the FM configuration is considered, the DOS peak at $E_F$ is largely reduced, and in particular, the DOS of the spin down channel nearly vanishes. Nevertheless, the energy gap in the spin up channel is filled, unlike in the case of c-CrGT. The –COHP also shows a much improved bonding situation as compared to the NM state. The total energy of the FM configurations is ~206 ± 11 meV/atom lower than the NM ones, and the average magnetic moment is 3.11 ± 0.49 $\mu_B$ per Cr atom (averaged over three amorphous models). Given the small but negative θ value obtained by fitting to the Curie–Weiss law, AFM coupling should come into play in a-GrGT. To model disordered magnetic configurations with FM and AFM coupling, we assigned positive magnetic moments to half of the Cr atoms, chosen in a random fashion, and negative moments to the remaining ones (denoted as "RM"). The DOS and –COHP profiles of these RM models also show a valley-like shape (Figure 6b), while slight deviations between the spin up and spin down channels are seen, owing to their highly disordered nature. We considered three RM states with different spin configurations for each amorphous model. Although it is difficult to obtain configurations with exactly zero total magnetization, we obtained a small net magnetization of ~0.06 $\mu_B$ per Cr atom averaged over the nine RM states. Besides, we also considered more complex RM states with 2/3 or 3/4 Cr atoms being assigned with positive magnetic moments (figure S8). The total energy of all RM states is ~202 ± 12 meV/atom lower than that of the NM states. Overall, the energy of the magnetically random states is within the error bar as compared to the FM amorphous states, which is in line with the experimental observation of spin-glass behavior in a-CrGT. No band gap was observed for a-CrGT models regardless of the type of magnetic configuration.

**2.6 Understanding the spin glass behavior.**

The FM order in c-CrGT has been attributed to a Cr−Te−Cr superexchange mechanism[48-51]. The foundational paper by Kanamori[63] showed how superexchange can lead to FM in magnetic compounds, such as crystalline CrGT, where the magnetic atoms form 90° bonds with the anions. The two relevant bonding mechanisms involve Te $p\sigma$ − Cr $de_g$ orbitals and Te $p\sigma$ − Cr $dt_{2g}$ orbitals. However, deviations of the Cr–Te–Cr bond angles from 90 degrees can undermine the ferromagnetic coupling. We qualitatively explain the coexistence of FM and AFM couplings in a-CrGT and thus its spin glass behavior, as follows. First, we examine the ADF of Te-centered motifs in more detail. The Cr−Te−Cr angle distribution in a-CrGT shows a major peak close to 90° and two additional peaks near 70° and 120° (**Figure 7**a). The major ADF peak corresponds to a pair of edge-sharing (*es*-) octahedra (two shared Te atoms; Figure 7b), which represents the local geometry in c-CrGT. Instead, the ADF peaks close to 120° and 70° correspond to pairs of corner-sharing (*cs*-) and face-sharing (*fs*-) octahedra, linked via one and three shared Te atoms, respectively. These motifs result in a variation in Cr–Cr interatomic distances in the amorphous phase: from the shortest in *fs*-octahedra ~3.0 Å to the longest in *cs*-octahedra above 4.6 Å (Figure 7c and Figure 6b). It is interesting to note that these Cr[Te$_6$] octahedra are the basic structural building blocks of three crystalline phases, namely, hexagonal (h-) CrTe, trigonal (1T-) CrTe$_2$ and monoclinic (m-) CrTe$_3$. The DFT-relaxed structures of the three crystals, their bond lengths and angles, as well as their partial DOS are shown in Figure 7 d–f.

All these chromium telluride crystals are intrinsic magnetic materials. The 1T-CrTe$_2$ crystal resembles c-CrGT most closely, since its Cr[Te$_6$] octahedra also take the *es*-configuration but without the Ge–Ge dimer (Figure 7e and Figure 1a). This layered metastable 1T-CrTe$_2$ vdW thin film has been synthesized experimentally recently[64-66], and has been demonstrated to be a ferromagnetic metal. Compared to the case of c-CrGT, the six Te atoms per atomic slab also form tetrahedral coordination environments with the two Ge atoms, and this additional orbital hybridization creates a small energy gap at $E_F$. The Cr atoms in h-CrTe[67] form *fs*-Cr[Te$_6$] octahedra along the *c* axis but *es*-Cr[Te$_6$] octahedra in the *ab* plane (Figure 7d), and this alloy has been proved to be a ferromagnetic metal[67, 68]. CrTe$_3$ is a vdW material, but its structure is more complex than that of 1T-CrTe$_2$, since each CrTe$_3$ layer consists of 8 Cr and 24 Te atoms, forming both *es*- and *cs*-Cr[Te$_6$] octahedra. This alloy shows AFM ordering and semiconducting behavior with a narrow band gap of 0.3 eV[69], and the antiferromagnetism stems from the distorted *es*-Cr[Te$_6$] octahedra (Cr−Te−Cr bond angle ~80º) forming lozenge-shaped tetramers[69, 70]. The four Cr-octahedra in a tetramer also share corners with those in neighboring tetramers (Cr−Te−Cr bond angle ~115º). Our DFT calculations show an energy difference between the AFM and FM configuration $E_{AFM} - E_{FM}$ as 8.9, 3.2, and −14.7 meV/atom for crystalline CrTe, CrTe$_2$ and CrTe$_3$, respectively, supporting the experimental observations[64, 68, 69]. The optimized lattice parameters, typical interatomic distances and bond angles are shown in Table 1.

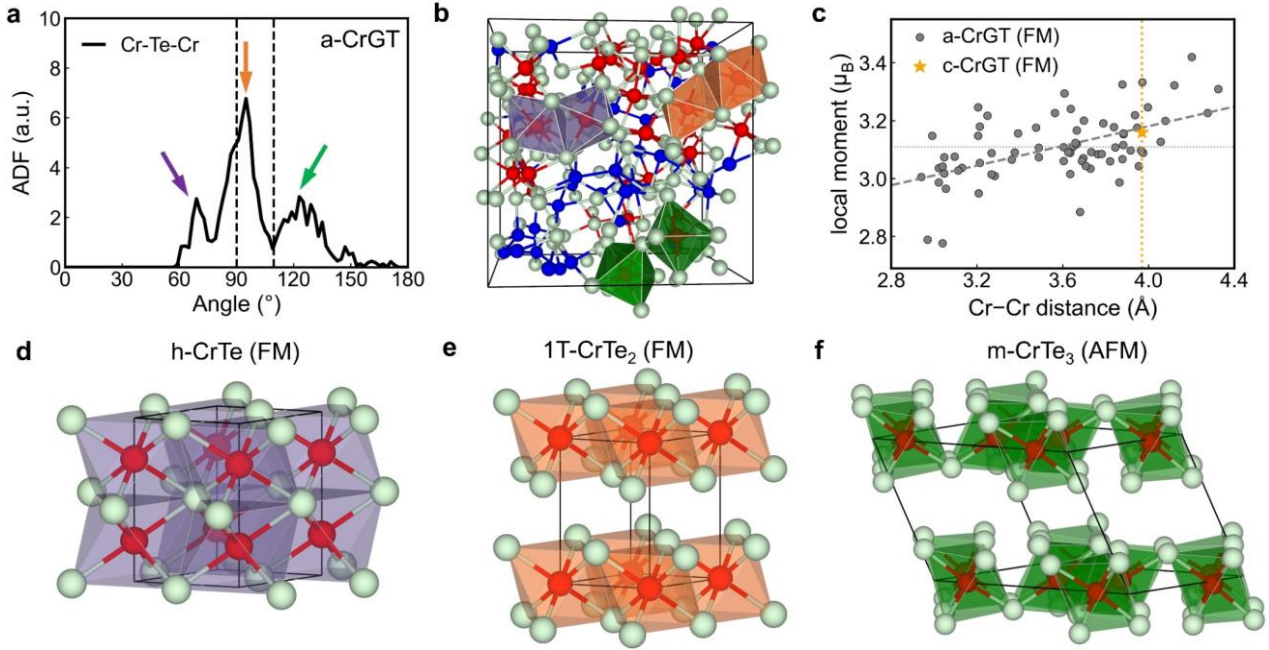

**Figure 7. Local structural motifs in a-CrGT and corresponding crystalline phases. a** The ADF of Cr−Te−Cr in a-CrGT. **b** One a-CrGT model with typical *fs*- (purple), *es*- (orange) and *cs*- (green) Cr[Te$_6$] octahedra being highlighted. **c** The local moments of Cr atoms within the Cr[Te$_6$] octahedra as a function of the distance with their nearest Cr atom for three a-CrGT models in FM configuration (grey dots). The c-CrGT model in FM configuration is shown for comparison (yellow point). **d-f** The DFT-relaxed crystalline structures and the corresponding partial DOS of h-CrTe, 1T-CrTe$_2$ and m-CrTe$_3$. The Cr, Ge and Te atoms are rendered as red, blue and green spheres.

In general, all the structural motifs of the three chromium telluride crystals can be found in a-CrGT, as highlighted in Figure 7b. Yet, given the stronger angular disorder and more distorted chemical environments, the magnetic structure of a-CrGT is more complex. In addition to superexchange-induced FM and AFM couplings, direct exchange between Cr atoms (and, possibly, also *p-d* exchange coupling due to the absence of an energy gap) could also affect the couplings. Furthermore, the Cr−Cr distance also affects the magnitude of the local moments. We computed three a-CrGT models in the FM configuration, and plotted the local magnetic moment of the center Cr atoms of the Cr[Te$_6$] octahedra with respect to the nearest Cr−Cr interatomic distance. As shown in Figure 7c, the local moment is reduced as the Cr−Cr direct interaction gets stronger at a shorter interatomic distance. Moreover, the defective octahedral motifs with five Te neighbors or "wrong" bonds (i.e., Cr–Ge bonds) also affect the spin ordering of Cr atoms.

In all magnetic configurations we considered above, the crystalline phase of CrGT always shows a narrow band gap, while there is no energy gap in the amorphous phase. This contrast could result in a sizable difference in carrier concentration, giving rise to the inverse resistance change upon phase transition[27]. If the two phases are computed in the non-magnetic state, the predicted band gap is closed in both[55]. Hence, the intrinsic magnetic features of CrGT account for its unconventional phase-change behavior. If spin polarization is considered for a-CrGT AIMD calculations at 300 K, all the Cr atoms still form octahedral motifs locally just like at low temperatures (figure S9), yielding a high structural similarity to c-CrGT, regardless of the transition to the paramagnetic state. The robustness of Cr octahedral motifs could be the microscopic origin of the rapid crystallization of a-CrGT, which can be accomplished in 30 ns using a voltage pulse of 1.6 V in memory devices[27]. Further experimental and theoretical investigations, e.g. using *in situ* atomic-scale structural characterization experiments[71-74] and molecular dynamics simulations via machine-learned interatomic potentials[75-78], are anticipated to provide an in-depth understanding of the crystallization kinetics of amorphous CrGT.

Before closing, we advocate the potential utility of the CrGT a-c phase change. In previous mPCM research, amorphous Fe (~7 at. %) doped GST was shown to be a ferromagnetic semiconductor rather than a spin glass[34]. The decrease in magnetic moment upon amorphization measured at ~5 K was attributed to the stronger saturation by shorter chemical bonds around Fe impurities in the amorphous phase[35]. The magnetic properties of CrGT phases, in particular the discoveries discussed in this paper, make it a superior candidate for mPCM applications. At temperatures below ~20 K, the amorphous phase is a spin glass with metallic character, while the crystalline phase is a ferromagnetic semiconductor. Between 20 and 64 K, the crystalline phase still shows strong FM response with a net magnetization of over 22 emu/g (Figure 1e), while that of the amorphous phase is reduced towards zero (below 0.05 emu/g, Figure 3d) given the transition to the paramagnetic state. The crystalline and amorphous phase can be regarded as magnetic "ON" and "OFF" state, respectively. The much wider magnetic contrast window in CrGT could also enable multilevel encoding of magnetic moments, analogous to the multilevel electrical or optical programming of PCM devices via iterative RESET operations (partial amorphization of a memory cell)[79-81]. Given the stoichiometric composition of CrGT, the tendency towards phase separation should be lower as compared to doped PCMs. Nevertheless, it is important to enhance the magnetic transition temperature of CrGT for practical use[18-20].

## 3. Conclusion

We have combined experiments and computations to reveal the magnetic properties of the CrGT phase-change alloy in both crystalline and amorphous phases. Upon amorphization, both the structural pattern and spin configuration undergo a transition to the glass state. In this disordered phase, the Cr atoms still form octahedral environments coordinated by Te atoms, despite the presence of bond distortions and defective neighbors. Given the high concentration of Cr atoms in $Cr_2Ge_2Te_6$ (20 at. %), the medium-range order between Cr-centered octahedral motifs must be considered. In the crystalline phase, all Cr octahedra are connected by edge-sharing with Cr−Te−Cr angles of ~90º, which stabilizes the FM ordering via superexchange. However, in the amorphous phase, the Cr−Te−Cr angles between Cr octahedra shows three major peaks at ~70°, ~90° and ~120° – corresponding to the local structural motifs in the crystalline phases of ferromagnetic CrTe, ferromagnetic $CrTe_2$ (and $Cr_2Ge_2Te_6$) and antiferromagnetic $CrTe_3$, respectively. The more complex medium-range order and structural distortions give rise to the spin glass behavior in amorphous CrGT. This difference in magnetic interaction also results in a major difference in carrier concentration, giving rise to the inverse resistivity contrast upon phase transition. In a very recent work, a hybrid photonic–electronic processing array has been implemented using GST, which utilized electrical pulse induced Joule heating for switching and the associated changes in optical transmission for data processing[88]. This in-memory computing platform has enabled highly efficient image processing with a superior contrast-to-noise ratio for a much improved computing accuracy. We envision that the concurrently tunable magnetic, electrical and optical properties associated with the a-c phase change via nanoseconds pulses in CrGT can be exploited for multifunctional device applications. We envision that the concurrently tunable magnetic, electrical and optical properties associated with the a-c phase change via nanoseconds pulses in CrGT can be exploited for multifunctional device applications.

## 4. Experimental Section

### 4.1 Material synthesis.

The CrGT single-crystal bulk sample was grown via chemical vapor transport (CVT). High purity elemental Cr (99.99%) powder, Ge (99.999%) powder, and Te slices (99.999%) were mixed in a molar ratio 2:2:30 (excess Te was used as flux), sealed in an evacuated quartz ampoule, heated at 750 °C for 200 hours, and then slowly cooled to 500 °C over 50 hours. A single-crystal sample, of millimeters in size, was obtained via a centrifugation process to remove the flux at the surface. The ~200 nm-thick CrGT films were deposited on a $SiO_2$/Si substrate at room temperature by sputtering a stoichiometric $Cr_2Ge_2Te_6$ target in high vacuum (base pressure of less than $\sim 2 \times 10^{-5}$ Pa). The deposition rate was set to 8.5 nm min$^{-1}$. A ~10 nm thick $SiO_2$ capping layer was grown on top of the CrGT film inside the vacuum chamber to prevent oxidation and evaporation.

### 4.2 Experimental characterization.

The Keithley 2636B source meter and the Instec mK200 hot stage with a temperature accuracy of 0.001 °C were used for the electrical measurements. The resistance of the film as a function of temperature (R–T) was measured in situ under an Ar atmosphere using a two-point probe method with a heating rate of 10 °C / min, where the probe electrode was tungsten. The structures of the as-deposited and post-annealed thin films were investigated by X-ray diffraction (XRD) with Cu $K_\alpha$ radiation (Bruker D8 ADVANCE). The transmission electron microscopy (TEM) specimens were prepared by a dual beam focused ion beam (FIB) system (Helios NanoLab 600i, FEI) with a Ga ion beam operated at 30 kV. The spherical aberration corrected (Cs-corrected) scanning transmission electron microscopy (STEM) - high angle annular dark field (HAADF) imaging experiments were performed on a JEOL ARM200F STEM with a probe aberration corrector, operated at 200 keV. The energy-dispersive X-ray (EDX) and bright-field spectroscopy experiments were performed on a Talos-F200X operated at 200 kV. The direct-current and alternating-current magnetic measurements were carried out using a superconducting quantum interference device (SQUID) magnetometer (Quantum Design MPMS3-VSM) with a maximum field of 7 Tesla. Raman spectra were collected using Renishaw inVia Qontor with a solid-state 532 nm laser for the excitation. The laser power was set as 0.25 mW, and an exposure time of 2 s with 50 cycles was used.

### 4.3 *Ab initio* calculations.

*Ab initio* molecular dynamics (AIMD) simulations based on density functional theory (DFT) were performed using the second-generation Car-Parrinello molecular dynamics scheme[82] implemented in CP2K[83], employing the Goedecker pseudopotentials, the Perdew–Burke–Ernzerhof functional and semi-empirical van der Waals corrections[84]. The self-consistent calculations in the CP2K code (QUICKSTEP) are based on a mixed Gaussian and plane-wave approach[83]. The Kohn–Sham orbitals were expanded in a Gaussian-type basis set with double-/triple-ζ polarization quality, whereas the charge density was expanded in plane waves, with a cutoff of 300 Ry. The AIMD calculations were carried out in the canonical (NVT) ensemble with a time step of 2 fs. Both spin-polarized and non-spin-polarized configurations were considered for the AIMD calculations. The spin-polarization was achieved using α and β orbitals, and no spin restriction was applied. Further structural relaxation was carried out using VASP[85] prior to the electronic structure calculations and magnetic property analyses. The projector augmented-wave (PAW) method and the PBE functional were used with an energy cutoff of 450 eV. The chemical bonding analyses were performed using LOBSTER[86]. The unit cell of the crystalline model contains 30 atoms, and each amorphous model contains 180 atoms. Three independent amorphous models were generated via melt-quenching. The atomic structures were visualized using VESTA[87].

### Data availability
Relevant data supporting the key findings of this study are available within the article and the Supplementary Information. All raw data are available from the corresponding author upon reasonable request.

### Competing Interests
The authors declare no competing interests.

### Acknowledgements

We acknowledge the Shenzhen 6Carbon Technology Co., Ltd. for providing the single crystal sample, and Dr. Yudong Cheng for the suggestion on powder sample preparation. We thank Qizhong Zhao for his technical support on magnetic measurement, and Jiao Li and Danli Zhang for their helps on TEM characterization. J.-J.W. thanks the support of National Natural Science Foundation of China (62204201). E.M. thanks the support of National Natural Science Foundation of China (52150710545). J.Z. thanks the support of National Natural Science Foundation of China (21903063). W.Z. thanks the support of 111 Project 2.0 (BP0618008). W.Z. and E.M. acknowledge support of XJTU for their work at CAID. The authors acknowledge the support of the International Joint Laboratory for Micro/Nano Manufacturing and Measurement Technologies and the HPC platform of Xi'an Jiaotong

**Supporting Information**
Supporting Information is available from the Wiley Online Library or from the author

**References**
[1]  M. Wuttig, N. Yamada, *Nat. Mater.* **2007**, *6*, 824.
[2]  W. Zhang, R. Mazzarello, M. Wuttig, E. Ma, *Nat. Rev. Mater.* **2019**, *4*, 150.
[3]  M. Xu, X. Mai, J. Lin, W. Zhang, Y. Li, Y. He, H. Tong, X. Hou, P. Zhou, X. Miao, *Adv. Funct. Mater.* **2020**, *30*, 2003419.
[4]  X.-B. Li, N.-K. Chen, X.-P. Wang, H.-B. Sun, *Adv. Funct. Mater.* **2018**, *28*, 1803380.
[5]  T.-T. Jiang, X.-D. Wang, J.-J. Wang, H.-Y. Zhang, L. Lu, C. Jia, M. Wuttig, R. Mazzarello, W. Zhang, E. Ma, *Fundam. Res.* **2022**, DOI: 10.1016/j.fmre.2022.09.010.
[6]  X. Chen, Y. Xue, Y. Sun, J. Shen, S. Song, M. Zhu, Z. Song, Z. Cheng, P. Zhou, *Adv. Mater.* **2022**, *34*, 2203909.
[7]  W. Zhang, H. Zhang, S. Sun, X. Wang, Z. Lu, X. Wang, J.-J. Wang, C. Jia, C. F. Schon, R. Mazzarello, E. Ma, M. Wuttig, *Adv. Sci.* **2023**, *10*, 2300901.
[8]  J. J. Wang, H. M. Zhang, X. D. Wang, L. Lu, C. Jia, W. Zhang, R. Mazzarello, *Adv. Mater. Technol.* **2022**, *7*, 2200214.
[9]  Z. Yang, B. Li, J.-J. Wang, X.-D. Wang, M. Xu, H. Tong, X. Cheng, L. Lu, C. Jia, M. Xu, X. Miao, W. Zhang, E. Ma, *Adv. Sci.* **2022**, *9*, 2103478.
[10] J.-J. Wang, X. Wang, Y. Cheng, J. Tan, C. Nie, Z. Yang, M. Xu, X. Miao, W. Zhang, E. Ma, *Mater. Futures* **2022**, *1*, 045302.
[11] F. Pan, S. Gao, C. Chen, C. Song, F. Zeng, *Mater. Sci. Eng. R-Rep* **2014**, *83*, 1.
[12] S. Bhatti, R. Sbiaa, A. Hirohata, H. Ohno, S. Fukami, S. N. Piramanayagam, *Mater. Today* **2017**, *20*, 530.
[13] C. Y. Wang, C. Wang, F. Meng, P. Wang, S. Wang, S. J. Liang, F. Miao, *Adv. Electron. Mater.* **2019**, *5*, 1901107.
[14] Z. Wang, H. Wu, G. W. Burr, C. S. Hwang, K. L. Wang, Q. Xia, J. J. Yang, *Nat. Rev. Mater.* **2020**, *5*, 173.
[15] C. Gong, L. Li, Z. Li, H. Ji, A. Stern, Y. Xia, T. Cao, W. Bao, C. Wang, Y. Wang, Z. Q. Qiu, R. J. Cava, S. G. Louie, J. Xia, X. Zhang, *Nature* **2017**, *546*, 265.
[16] Z. Wang, T. Zhang, M. Ding, B. Dong, Y. Li, M. Chen, X. Li, J. Huang, H. Wang, X. Zhao, Y. Li, D. Li, C. Jia, L. Sun, H. Guo, Y. Ye, D. Sun, Y. Chen, T. Yang, J. Zhang, S. Ono, Z. Han, Z. Zhang, *Nat. Nanotechnol.* **2018**, *13*, 554.
[17] V. Ostwal, T. Shen, J. Appenzeller, *Adv. Mater.* **2020**, *32*, 1906021.
[18] I. A. Verzhbitskiy, H. Kurebayashi, H. Cheng, J. Zhou, S. Khan, Y. P. Feng, G. Eda, *Nat. Electron.* **2020**, *3*, 460.
[19] W. Zhuo, B. Lei, S. Wu, F. Yu, C. Zhu, J. Cui, Z. Sun, D. Ma, M. Shi, H. Wang, W. Wang, T. Wu, J. Ying, S. Wu, Z. Wang, X. Chen, *Adv. Mater.* **2021**, *33*, 2008586.
[20] D. Bhoi, J. Gouchi, N. Hiraoka, Y. Zhang, N. Ogita, T. Hasegawa, K. Kitagawa, H. Takagi, K. H. Kim, Y. Uwatoko, *Phys. Rev. Lett.* **2021**, *127*, 217203.
[21] W. Ge, K. Xu, W. Xia, Z. Yu, H. Wang, X. Liu, J. Zhao, X. Wang, N. Yu, Z. Zou, Z. Yan, L. Wang, M. Xu, Y. Guo, *J. Alloys Compd.* **2020**, *819*, 153368.
[22] W. Cai, L. Yan, S. K. Chong, J. Xu, D. Zhang, V. V. Deshpande, L. Zhou, S. Deemyad, *Phys. Rev. B* **2022**, *106*, 085116.
[23] D. Yang, W. Yao, Q. Chen, K. Peng, P. Jiang, X. Lu, C. Uher, T. Yang, G. Wang, X. Zhou, *Chem. Mater.* **2016**, *28*, 1611.
[24] L. Xie, L. Guo, W. Yu, T. Kang, R.-K. Zheng, K. Zhang, *Nanotechnology* **2018**, *29*, 464002.
[25] S. Xing, J. Zhou, X. Zhang, S. Elliott, Z. Sun, *Prog. Mater. Sci.* **2023**, *132*, 101036.
[26] S. Hatayama, T. Yamamoto, S. Mori, Y.-H. Song, Y. Sutou, *ACS Appl. Mater. Interfaces* **2022**, *14*, 44604.

[27] S. Hatayama, Y. Sutou, S. Shindo, Y. Saito, Y. H. Song, D. Ando, J. Koike, *ACS Appl. Mater. Interfaces* **2018**, *10*, 2725.
[28] S. Hatayama, Y.-H. Song, Y. Sutou, *Mater. Sci. Semicond. Process.* **2021**, *133*, 105961.
[29] S. Hatayama, D. Ando, Y. Sutou, *J. Phys. D: Appl. Phys.* **2019**, *52*, 325111.
[30] Q. Wang, B. Liu, Y. Xia, Y. Zheng, R. Huo, Q. Zhang, S. Song, Y. Cheng, Z. Song, S. Feng, *Appl. Phys. Lett.* **2015**, *107*, 222101.
[31] Q. Wang, P. M. Konze, G. Liu, B. Liu, X. Chen, Z. Song, R. Dronskowski, M. Zhu, *J. Phys. Chem. C* **2019**, *123*, 30640.
[32] Q. Wang, B. Liu, Y. Xia, Y. Zheng, R. Huo, M. Zhu, S. Song, S. Lv, Y. Cheng, Z. Song, S. Feng, *Phys. Status Solidi RRL* **2015**, *9*, 470.
[33] J. Hu, C. Lin, L. Peng, T. Wei, W. Li, Y. Ling, Q. Liu, M. Cheng, S. Song, Z. Song, J. Zhou, Y. Cheng, Y. Zheng, Z. Sun, B. Liu, *J. Alloys Compd.* **2022**, *908*, 164593.
[34] W.-D. Song, L.-P. Shi, X.-S. Miao, C.-T. Chong, *Adv. Mater.* **2008**, *20*, 2394.
[35] Y. Li, R. Mazzarello, *Adv. Mater.* **2012**, *24*, 1429.
[36] W. Zhang, I. Ronneberger, Y. Li, R. Mazzarello, *Adv. Mater.* **2012**, *24*, 4387.
[37] W. Zhang, I. Ronneberger, Y. Li, R. Mazzarello, *Sci. Adv. Mater.* **2014**, *6*, 1655.
[38] T. Fukushima, H. Katayama-Yoshida, K. Sato, H. Fujii, E. Rabel, R. Zeller, P. H. Dederichs, W. Zhang, R. Mazzarello, *Phys. Rev. B* **2014**, *90*, 144417.
[39] M. Mézard, G. Parisi, M. A. Virasoro, *Spin glass theory and beyond: An Introduction to the Replica Method and Its Applications*, World Scientific Publishing Company, 1987.
[40] T. Egami, *Rep. Prog. Phys.* **1984**, *47* 1601.
[41] R. Hasegawa, *J. Magn. Magn. Mater.* **1991**, *100*, 1.
[42] P. Marín, A. Hernando, *J. Magn. Magn. Mater.* **2000**, *215-216*, 729.
[43] D. Azuma, N. Ito, M. Ohta, *J. Magn. Magn. Mater.* **2020**, *501*, 166373.
[44] W. Liu, H. Zhang, J. A. Shi, Z. Wang, C. Song, X. Wang, S. Lu, X. Zhou, L. Gu, D. V. Louzguine-Luzgin, M. Chen, K. Yao, N. Chen, *Nat. Commun.* **2016**, *7*, 13497.
[45] Q. Li, R. Qiao, A. Mehta, W. Lü, T. Zhou, E. Arenholz, C. Wang, Y. Chen, L. Li, Y. Tian, L. Bai, Z. Hussain, R. Zheng, W. Yang, S. Yan, *Sci. Bull.* **2020**, *65*, 1718.
[46] S. Yin, C. Xiong, C. Chen, X. Zhang, *Phys. Chem. Chem. Phys.* **2020**, *22*, 8672.
[47] Y. Sun, R. C. Xiao, G. T. Lin, R. R. Zhang, L. S. Ling, Z. W. Ma, X. Luo, W. J. Lu, Y. P. Sun, Z. G. Sheng, *Appl. Phys. Lett.* **2018**, *112*, 072409.
[48] V. Carteaux, D. Brunet, G. Ouvrard, G. Andre, *J. Phys. Condens. Matter* **1995**, *7*, 69.
[49] Y. F. Li, W. Wang, W. Guo, C. Y. Gu, H. Y. Sun, L. He, J. Zhou, Z. B. Gu, Y. F. Nie, X. Q. Pan, *Phys. Rev. B* **2018**, *98*, 125127.
[50] Z. Hao, H. Li, S. Zhang, X. Li, G. Lin, X. Luo, Y. Sun, Z. Liu, Y. Wang, *Sci. Bull.* **2018**, *63*, 825.
[51] J. Li, J. Feng, P. Wang, E. Kan, H. Xiang, *Sci. China Phys. Mech. Astron.* **2021**, *64*, 286811.
[52] Z. Q. Qiu, S. D. Bader, *Rev. Sci. Instrum.* **2000**, *71*, 1243
[53] S.-X. Peng, Y. Cheng, J. Pries, S. Wei, H.-B. Yu, M. Wuttig, *Sci. Adv.* **2020**, *6*, eaay6726.
[54] J. A. Mydosh, *Spin Glasses: An Experimental Introduction*, Taylor & Francis, London 1993.
[55] M. Xu, Y. Guo, Z. Yu, K. Xu, C. Chen, H. Tong, X. Cheng, M. Xu, S. Wang, C. Z. Wang, K.-M. Ho, X. Miao, *J. Mater. Chem. C* **2019**, *7*, 9025.
[56] S. Hatayama, Y. Shuang, P. Fons, Y. Saito, A. V. Kolobov, K. Kobayashi, S. Shindo, D. Ando, Y. Sutou, *ACS Appl. Mater. Interfaces* **2019**, *11*, 43320.
[57] V. L. Deringer, W. Zhang, M. Lumeij, S. Maintz, M. Wuttig, R. Mazzarello, R. Dronskowski, *Angew. Chem. Int. Ed.*

**2014**, *53*, 10817.

[58] F. C. Mocanu, K. Konstantinou, J. Mavračić, S. R. Elliott, *Phys. Status Solidi RRL* **2021**, *15*, 2000485.

[59] R. Dronskowski, P. E. Blöchl, *J. Phys. Chem.* **1993**, *97*, 8617.

[60] G. A. Landrum, R. Dronskowski, *Angew. Chem. Int. Ed.* **2000**, *39*, 1560.

[61] J. Brgoch, C. Goerens, B. P. Fokwa, G. J. Miller, *J. Am. Chem. Soc.* **2011**, *133*, 6832.

[62] B. P. Fokwa, H. Lueken, R. Dronskowski, *Chem. Eur. J.* **2007**, *13*, 6040.

[63] J. Kanamori, *J. Phys. Chem. Solids* **1959**, *10*, 87.

[64] D. C. Freitas, R. Weht, A. Sulpice, G. Remenyi, P. Strobel, F. Gay, J. Marcus, M. Nunez-Regueiro, *J. Phys. Condens. Matter.* **2015**, *27*, 176002.

[65] X. Sun, W. Li, X. Wang, Q. Sui, T. Zhang, Z. Wang, L. Liu, D. Li, S. Feng, S. Zhong, H. Wang, V. Bouchiat, M. Nunez Regueiro, N. Rougemaille, J. Coraux, A. Purbawati, A. Hadj-Azzem, Z. Wang, B. Dong, X. Wu, T. Yang, G. Yu, B. Wang, Z. Han, X. Han, Z. Zhang, *Nano Res.* **2020**, *13*, 3358.

[66] X. Zhang, Q. Lu, W. Liu, W. Niu, J. Sun, J. Cook, M. Vaninger, P. F. Miceli, D. J. Singh, S. W. Lian, T. R. Chang, X. He, J. Du, L. He, R. Zhang, G. Bian, Y. Xu, *Nat. Commun.* **2021**, *12*, 2492.

[67] J. Dijkstra, H. H. Weitering, C. F. v. Bruggen, C. Haas, R. A. d. Groot, *J. Phys. Conden. Matt.* **1989**, *1*, 9141.

[68] H. Wu, W. Zhang, L. Yang, J. Wang, J. Li, L. Li, Y. Gao, L. Zhang, J. Du, H. Shu, H. Chang, *Nat. Commun.* **2021**, *12*, 5688.

[69] M. A. McGuire, V. O. Garlea, S. Kc, V. R. Cooper, J. Yan, H. Cao, B. C. Sales, *Phys. Rev. B* **2017**, *95*, 144421.

[70] J. Yao, H. Wang, B. Yuan, Z. Hu, C. Wu, A. Zhao, *Adv. Mater.* **2022**, *34*, 2200236.

[71] K. Ding, J. Wang, Y. Zhou, H. Tian, L. Lu, R. Mazzarello, C. Jia, W. Zhang, F. Rao, E. Ma, *Science* **2019**, *366*, 210.

[72] T.-T. Jiang, J.-J. Wang, L. Lu, C.-S. Ma, D.-L. Zhang, F. Rao, C.-L. Jia, W. Zhang, *APL Mater.* **2019**, *7*, 081121.

[73] T.-T. Jiang, X.-D. Wang, J.-J. Wang, Y.-X. Zhou, D.-L. Zhang, L. Lu, C.-L. Jia, M. Wuttig, R. Mazzarello, W. Zhang, *Acta Mater.* **2020**, *187*, 103.

[74] X. Wang, K. Ding, M. Shi, J. Li, B. Chen, M. Xia, J. Liu, Y. Wang, J. Li, E. Ma, Z. Zhang, H. Tian, F. Rao, *Mater. Today* **2022**, *54*, 52.

[75] G. C. Sosso, G. Miceli, S. Caravati, F. Giberti, J. Behler, M. Bernasconi, *J. Phys. Chem. Lett.* **2013**, *4*, 4241.

[76] F. C. Mocanu, K. Konstantinou, T. H. Lee, N. Bernstein, V. L. Deringer, G. Csányi, S. R. Elliott, *J. Phys. Chem. B* **2018**, *122*, 8998.

[77] V. L. Deringer, M. A. Caro, G. Csányi, *Adv. Mater.* **2019**, *31*, 1902765.

[78] Y. Zhou, W. Zhang, E. Ma, V. L. Deringer, *arxiv:2207.14228* **2022**.

[79] A. Sebastian, M. Le Gallo, G. W. Burr, S. Kim, M. BrightSky, E. Eleftheriou, *J. Appl. Phys.* **2018**, *124*, 111101.

[80] W. Zhang, R. Mazzarello, E. Ma, *MRS Bulletin* **2019**, *44*, 686.

[81] B. J. Shastri, A. N. Tait, T. Ferreira de Lima, W. H. P. Pernice, H. Bhaskaran, C. D. Wright, P. R. Prucnal, *Nat. Photon.* **2021**, *15*, 102.

[82] T. Kühne, M. Krack, F. Mohamed, M. Parrinello, *Phys. Rev. Lett.* **2007**, *98*, 066401.

[83] J. Hutter, M. Iannuzzi, F. Schiffmann, J. VandeVondele, *WIREs Comput. Mol. Sci.* **2014**, *4*, 15.

[84] S. Grimme, J. Antony, S. Ehrlich, H. Krieg, *J. Chem. Phys.* **2010**, *132*, 154104

[85] G. Kresse, D. Joubert, *Phys. Rev. B* **1999**, *59*, 1758.

[86] R. Nelson, C. Ertural, J. George, V. L. Deringer, G. Hautier, R. Dronskowski, *J. Comput. Chem.* **2020**, *41*, 1931.

[87] K. Momma, F. Izumi, *J. Appl. Cryst.* **2011**, *44*, 1272.

[88] W. Zhou, B. Dong, N. Farmakidis, X. Li, N. Youngblood, K. Huang, Y. He, C. David Wright, W. Pernice, H. Bhaskaran, *Nat. Commun.* **2023**, *14*, 2887.

# Supporting Information

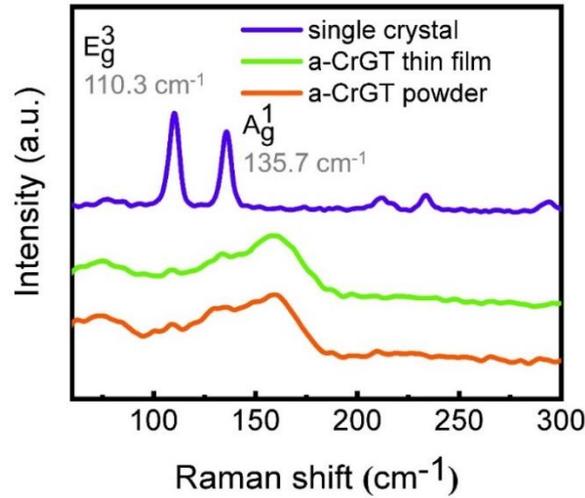

**Figure S1. Raman spectroscopy experiments.** In the CrGT single crystal, two primary modes are found at $E_g^3$ ~110.3 cm$^{-1}$ and $A_g^1$ ~135.7 cm$^{-1}$. Two small humps are also visible in the two amorphous CrGT samples at these two frequencies. A primary peak is observed at ~158.6 cm$^{-1}$ for both amorphous samples, and the two curves are overall very similar.

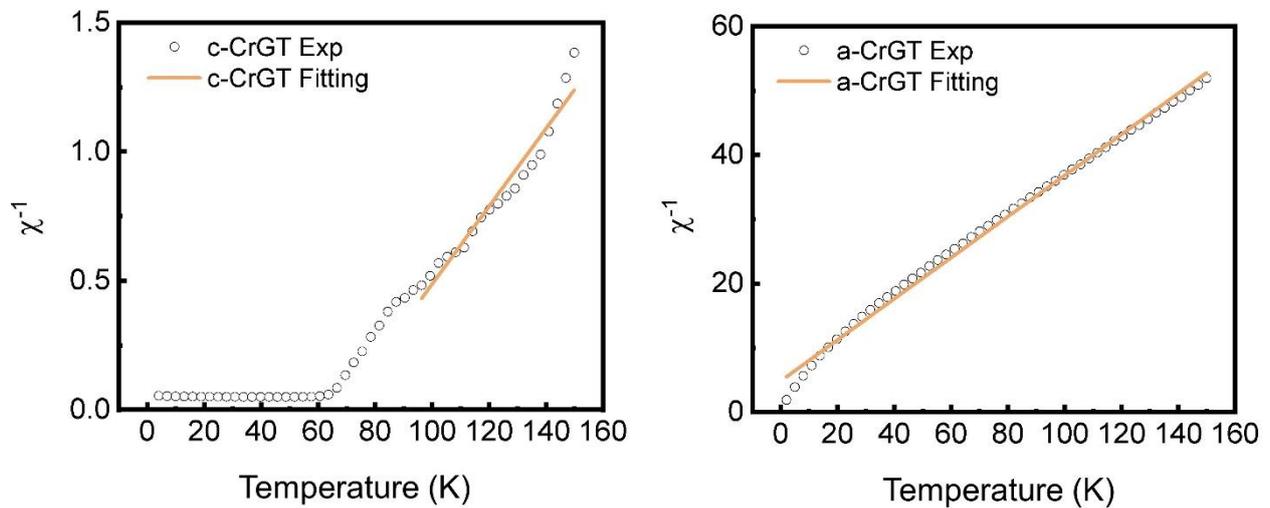

**Figure S2. Magnetic analysis.** The inverse susceptibility $\chi^{-1}(T)$ measured under 2000 Oe in ZFC. The curves measured for the single-crystal sample and amorphous powder sample are shown in the left and right panel, respectively, and are fitted by the Curie-Weiss law $\chi^{-1} = \frac{T-\theta}{c}$. Following Ref. [48], the single-crystal sample data above 100 K are used for linear fitting, and the obtained θ value equals 67.68 K, consistent with literature data. Regarding the amorphous sample, the data generally follow a linear change, and the obtained θ value equals –15.22 K.

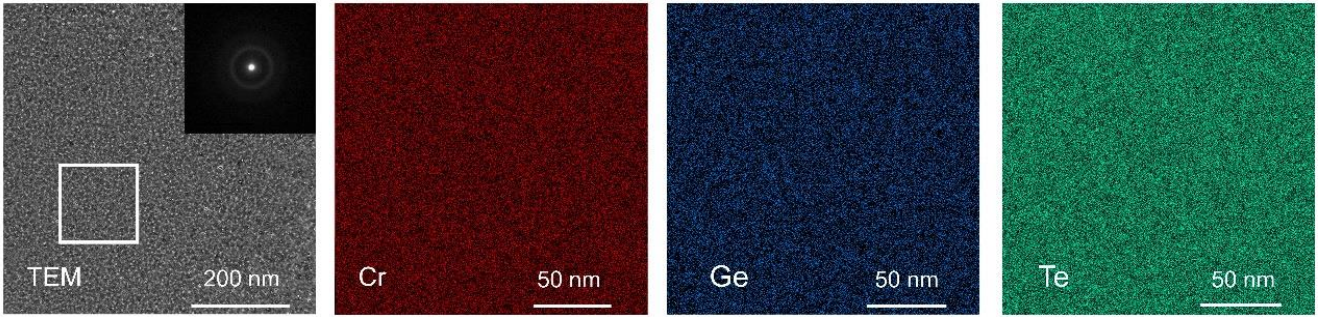

**Figure S3. TEM characterization of as-deposited CrGT thin films.** The bright-field TEM images and the corresponding SAED patterns with halo rings show the amorphous nature of the as-deposited CrGT films. The corresponding EDX mapping indicates that the distribution of Cr, Ge and Te atoms is homogenous at the length scale of tens of nanometers.

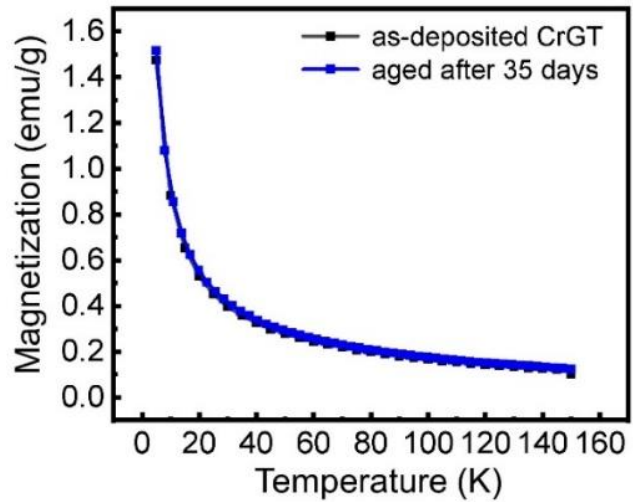

**Figure S4. Magnetic measurements of a-CrGT before and after aging.** The M-T curves measured for the a-CrGT powder sample before and after aging (after 35 days) are nearly identical. The applied magnetic field is 10 kOe.

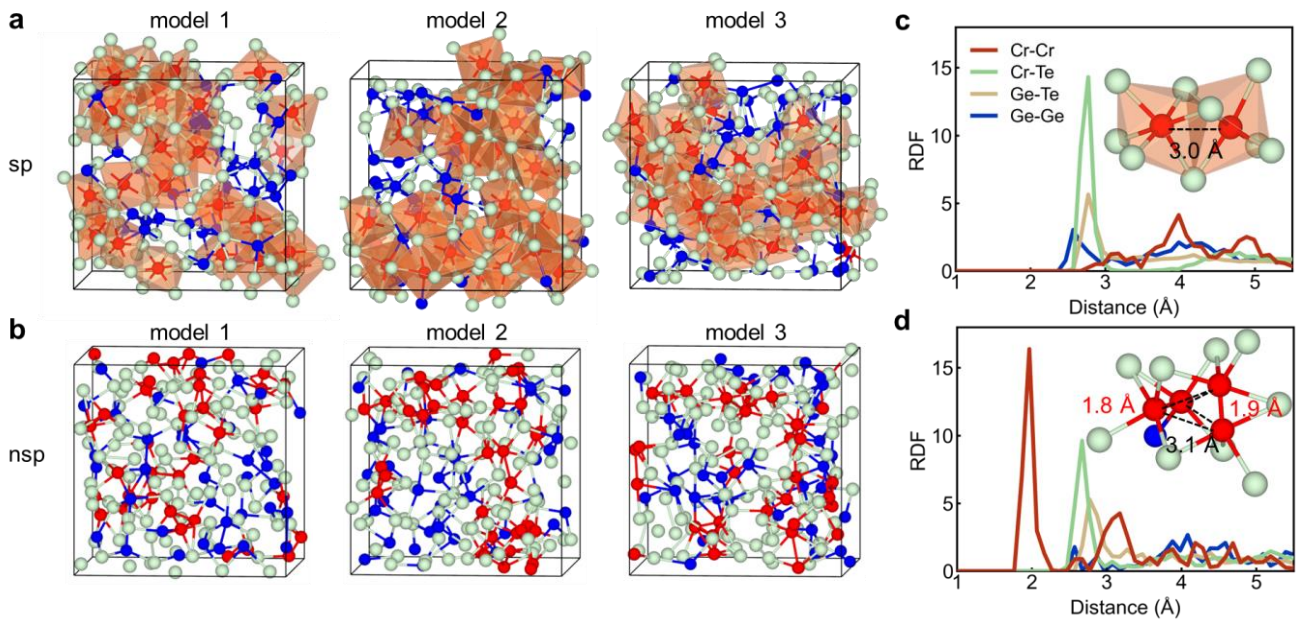

**Figure S5. Snapshots and RDF curves of spin-polarized (sp) and non-spin-polarized (nsp) a-CrGT. a,b** Snapshots of sp- (a) and nsp- (b) a-CrGT models. **c,d** The RDF curves of sp- (c) and nsp- (d) a-CrGT models annealing at 10K for 30ps with insets showing typical Cr local patterns. Cr, Ge and Te atoms are rendered as red, blue, and green spheres, respectively. Cr-centered octahedra are highlighted in orange. The sp-model 1 and the sp-RDF curves are shown in Fig. 5 in the main text.

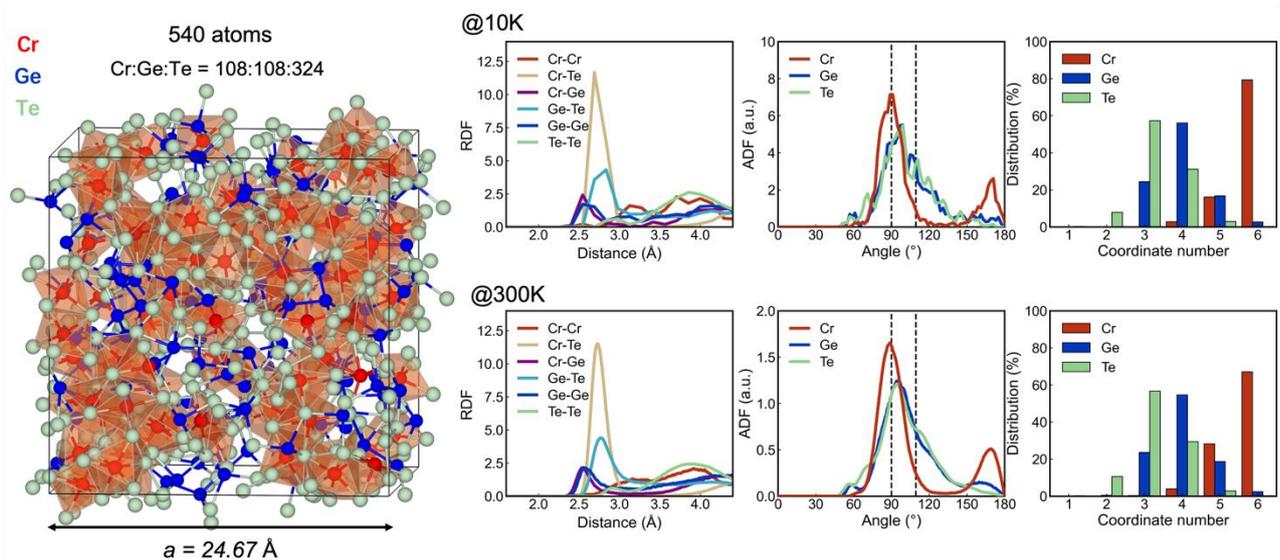

**Figure S6. A large amorphous CrGT model of 540 atoms with spin polarization.** The snapshot of the model together with RDF, ADF and CN distribution. The structural features of this big model are consistent with those of the small models shown in the main text.

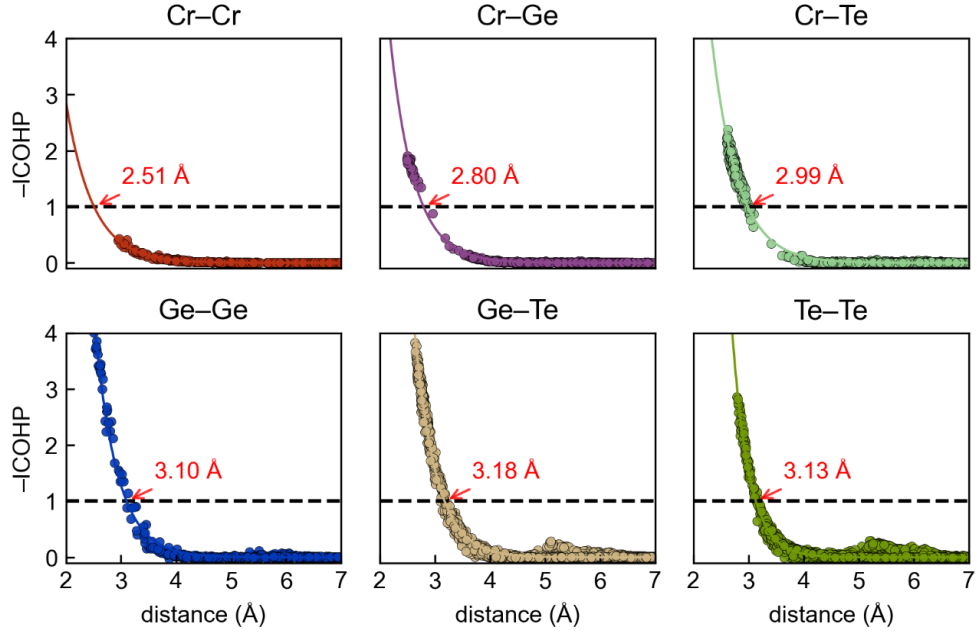

**Figure S7. Cutoffs for interatomic distance determined using the −COHP integral.** Following Ref. [58], the cutoff values for amorphous CrGT were determined as Cr−Cr 2.51 Å, Cr−Ge 2.80 Å, Cr−Te 2.99 Å, Ge−Ge 3.10 Å, Ge−Te 3.18 Å and Te−Te 3.13 Å.

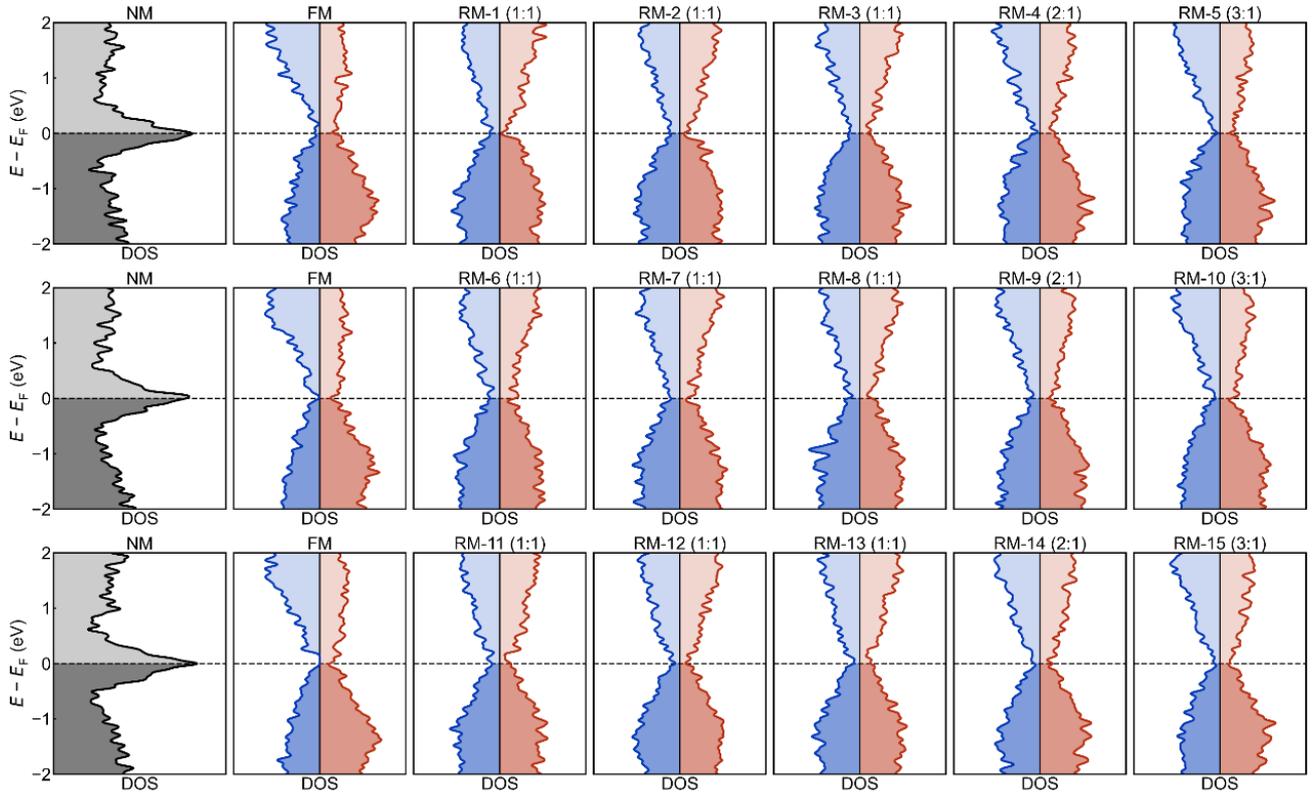

**Figure S8. Electronic structure of spin-polarized a-CrGT models in NM, FM and RM configurations.** For RM, finite magnetic moments were assigned to 1/2, 2/3 or 3/4 of the Cr atoms with positive values and the rest of them with negative values in a random fashion, denoted as (1:1), (2:1) and (3:1) in the corresponding panels. In total, 15 RM configurations were considered using 3 amorphous models.

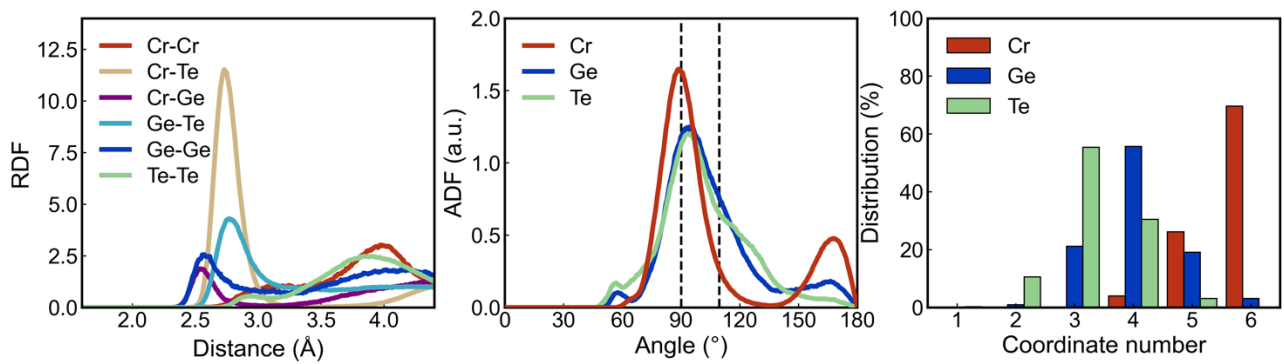

**Figure S9. Structural analysis of spin-polarized a-CrGT models annealed at 300 K.** The structural data were collected at 300K over 30 ps over three independent melt-quenched amorphous models.